\documentclass[aps,prx,longbibliography,a4paper,notitlepage,reprint,nofootinbib
]{revtex4-2}

\usepackage{amsmath,amssymb,amsfonts} % standard AMS packages
\usepackage{mathrsfs} % use \mathscr{} for script letters in math
\usepackage{mathtools} % for proper typesetting of := and =:
\usepackage{color}
\usepackage[usenames,dvipsnames]{xcolor}
\usepackage{graphicx,float}
\usepackage{hyperref}
\hypersetup{colorlinks,linkcolor=blue,citecolor=blue,urlcolor=blue}
\usepackage{cleveref}
\usepackage{soul}
\usepackage{siunitx}
\usepackage[inline]{enumitem}
\usepackage{titletoc}
\usepackage[switch]{lineno} 
\usepackage{lipsum}

\makeatletter
\newcommand{\fmarki}{*}
\newcommand{\fmarkii}{\ensuremath{\dagger}}
\def\@fnsymbol#1{{\ifcase#1\or \fmarki\or \fmarkii \else\@ctrerr\fi}}
\makeatother

\renewcommand{\fmarki}{$\dagger$}
\renewcommand{\fmarkii}{$\ddagger$}

\begin{document}
\title{
%All-optical single-shot superconducting qubit readout
%All-optical superconducting qubit readout
%Single-shot superconducting qubit readout via radio-over-fiber
%All-optical superconducting qubit readout via cryogenic radio-over-fiber
%All-optical qubit readout via cryogenic RF-photonics
%All-optical qubit readout via cryogenic radio-over-fiber
%All-optical qubit readout via radio-over-fiber at millikelvin temperature
%Radio over fiber superconducting qubit readout
%All-optical readout of a superconducting qubit via radio-over-fiber
%All-optical dispersive readout of a superconducting qubit 
%All-optical readout of a superconducting qubit using millikelvin radio over fiber
All-optical single-shot readout of a
superconducting qubit 
%readout
%Photonic superconducting qubit readout
}
\author{Georg Arnold$^\star$}
\email{georg.arnold@ist.ac.at}
\author{Thomas Werner$^\star$}
\author{Rishabh Sahu}
\author{Lucky N. Kapoor}
\author{Liu Qiu}
%\author{Liu Qiu$^\star$}
%\author{Rishabh Sahu$
%\author{William Hease}
\author{Johannes M. Fink}
\email{jfink@ist.ac.at}
\affiliation{Institute of Science and Technology Austria, Am Campus 1, 3400 Klosterneuburg, Austria}

\date{\today}

\begin{abstract}
\noindent
%\textbf{
%\textcolor{blue}{To be written}
%\textbf{
The rapid development of superconducting quantum hardware is expected to run into significant I/O restrictions due to the need for large-scale %and real-time 
%and modular 
error correction in a cryogenic 
%to be implemented in an ultra-cold 
environment~\cite{bravyi2022, beverland2022}.
%hoefler2023}. 
Classical data centers rely on fiber-optic interconnects to remove similar networking bottlenecks and to allow for reconfigurable, software-defined infrastructures~\cite{Cheng2018}. 
% This photonic approach has proven challenging to implement  .... bla bla state the problem and the most important results if a nature abstract
In the same spirit, 
ultra-cold electro-optic links~\cite{han2021} have been proposed~\cite{Joshi2022} and used to generate qubit control signals~\cite{lecocq2021}, or to replace cryogenic readout electronics. So far, the latter suffered from either low efficiency~\cite{youssefi2021}, low bandwidth and the need for additional microwave drives~\cite{delaney2022}, or breaking of Cooper pairs and qubit states~\cite{mirhosseini2020}. 
In this work we realize electro-optic microwave photonics at millikelvin temperatures to implement a radio-over-fiber qubit readout that does not require \textit{any} active or passive cryogenic microwave equipment. We demonstrate %high-fidelity, 
all-optical single-shot-readout 
%fidelities of $>$80\% 
by means of the Jaynes-Cummings nonlinearity~\cite{Reed2010a} in a 
%reflective and 
circulator-free readout scheme. 
%Even though the active optical heat load prevents scaling-up of this first generation implementation -  \LQ{$\leftarrow$
%maybe remove this?}
%surprisingly, %and based on high-fidelity quantum-non-demolition measurements, 
Importantly, we do not observe any %substantial 
direct radiation impact 
%of the telecom wavelength photons 
on the qubit state as verified with high-fidelity quantum-non-demolition measurements despite the absence of shielding elements.
This compatibility between superconducting circuits and telecom wavelength light is not only a prerequisite to establish modular quantum networks~\cite{ang2022}, it is also relevant for multiplexed readout of superconducting photon detectors~\cite{deCea2020} 
%\textcolor{blue}{maybe also a readout multiplexed, Johannes Heinsoo et al 2018 or Yu Chen et al 2012} 
and classical superconducting logic \cite{shen2023}. 
%[RSFQ paper from Hong]
%and multi-pixel . 
Moreover, this experiment showcases the potential of electro-optic radiometry in harsh environments - 
%n electronics-free RF 
%a photonic 
an electronics-free sensing principle that extends into the THz regime with applications in radio astronomy, planetary missions and earth observation~\cite{SantamariaBotello2018}.
\end{abstract}

\maketitle
%\section{Introduction}
\def\thefootnote{$\star$}
\footnotetext{These authors contributed equally to this work.}
\def\thefootnote{\arabic{footnote}}

%\section{Introduction}
The increasing demand 
%in modern communication 
for higher data transfer rates and energy efficiency alike has set the path to replacing electrical components by their optical counterparts. This is because of the substantially larger bandwidth of optical signals and the exceptionally low transmission loss in 
%optical 
fibers at telecom wavelengths. Recently, this transition affects not only long-distance communication but also short-range links
%data transfer 
within data centers~\cite{Cheng2018} or even on a single chip~\cite{Sun2015}. 
Moving the processors into a cryogenic environment can  
%Additionally, computing in cryogenic environments can 
decrease the power consumption of computation even further~\cite{Holmes2013}, increase the sensitivity of detection systems~\cite{deCea2020}, and interface classical control systems with cryogenic quantum processors directly~\cite{pauka_cryogenic_2021}.
%[(\textcolor{blue}{cite maybe more from supercable https://www.iarpa.gov/research-programs/supercables})], 
%This is relevant both for classical computing [cite] as well as for classical control, feedback and error correction of quantum processors [cite]. 
However, such an approach is also susceptible to transmission losses and related heating in electrical wires
%. Thus, it 
and thus might also benefit from a suitable, low-loss and low thermal conductivity optical~\cite{shen2023} or contactless~\cite{wang2023} links. 
%\LQ{Maybe combine first and second para.?}
%between room temperature 
%interconnects 
%and the cryogenic 
%processor 
%environment 
% [supercables]. 

\begin{figure}[t]
\includegraphics[width=\linewidth]{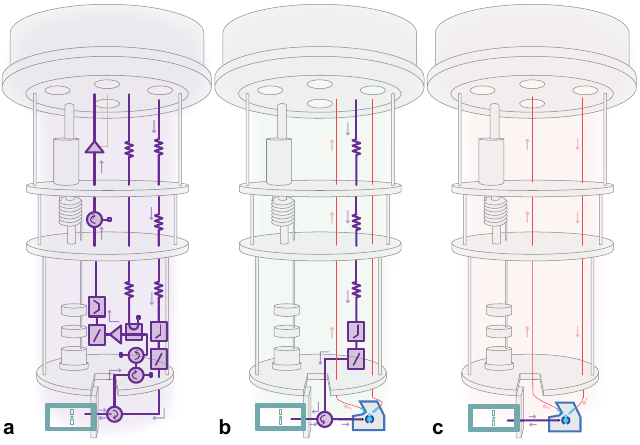}
\caption{
\textbf{Comparison of conventional 
%microwave 
and optical 
%superconducting 
qubit readout setups in a dilution refrigerator.} 
%Comparison of a conventional microwave readout and the heavily reduced all-optical readout scheme} 
\textbf{a}, microwave in - microwave out: Typical setup consisting of carefully thermalized 
%and high-resistivity (or superconducting) 
coaxial cables, attenuators, filters, circulators, a directional coupler, a driven parametric amplifier, and a dc-biased high-electron-mobility-transistor amplifier, all of which are approximately wavelength sized (centimeters).
\textbf{b}, microwave in - optics out: Reduced readout setup replacing the active microwave output 
%line 
components with an optically driven, resonant electro-optic transceiver.
\textbf{c}, optics in - optics out: all-optical, circulator-free %reflective 
qubit readout based on simultaneous microwave down- and up-conversion
%demodulation and modulation 
of an optical carrier.
%\LQ{This is confusing.} 
Here, all cryogenic microwave components are replaced by 
%2 optical fibers and 
a single electro-optic transceiver. 
}
%Schematic representation of the cavity electro-optic device.}
  %\LQ{Maybe a different color for the EO device and also in Fig2. Just to highlight the modularity.}
  %GA We should make the optical fibers more prominent (darker color, they are hard to see,
  % Label cQED system and EO converter so that the attention is immediately drawn to them)
  \label{fig:1}
\end{figure}

Quantum processors, such as superconducting platforms that operate at ultra-low temperatures of a few millikelvin, have particularly demanding I/O requirements. In stark contrast to classical processors, herein the number of external control and readout lines scales linearly with the number of qubits. 
%On the other hand, quantum computing propose a revolution in information and data processing \cite{Harrow2017}. Leading technologies for quantum processors, however, often need even colder environments than cryogenic classical computing, namely temperatures below 100 mK to avoid thermally initialized transitions of the quantum bit (qubit) states. Consequently, heating and transport losses are fundamental obstacles for the latter two technologies and their scalability to larger units and networks required to unfold the supreme potential. In more detail, superconducting qubits are individually addressed by a microwave signal guided through coaxial cables from room temperature to the cryogenic environment of a dilution refrigerator with a base temperature typically below 20 mK. In order to provide the required elimination of thermal noise from ambient environment a microwave signal is heavily attenuated by an order of 60 dB at different temperature stages leading to a great majority of the signal power being dissipated - in addition to the passive heat load due to the significant cable loss. 
Currently, the most powerful quantum processors utilize more than 100 qubits
%[cite most recent most fancy experiment], 
%\cite{arute2019, Zhu2022}, 
requiring hundreds of high-bandwidth coaxial cables with appropriate signal conditioning~\cite{kim_evidence_2023}, i.e. attenuation and careful thermalization on the input as well as isolation and low noise amplification on the output, see Fig.~\ref{fig:1}a. 
%for full control and readout capabilities. 
Considering the limited cooling power of dilution refrigerators, this architecture might allow for thousands of qubits~\cite{Krinner2019} given that advanced multiplexing strategies are employed~\cite{Chen2012, Heinsoo2018, acharya2023}. Thich is - ignoring space and financial constraints - still orders of magnitudes beneath the millions of qubits expected to be required for fault-tolerant universal quantum computing~\cite{Gidney2021, bravyi2022, hoefler2023}. 

Searching for ways to overcome these barriers, photonic links~\cite{youssefi2021,lecocq2021,Joshi2022} were identified as a promising alternative to conventional~\cite{Krinner2019}, cryo-CMOS~\cite{pauka_cryogenic_2021} or single flux quantum control~\cite{Liu2023_SFQ} of cryogenic quantum computing platforms. 
%[other approaches: McDermont, CMOS at cryo temps] In contrast to electrical signals in coaxial cables, microwave signals modulated on telecom waves are negligibly corrupted by thermal noise due to the high energy of the carrier wave, neither at cryogenic nor ambient environment. Thus, those signals are freed from the necessity of active attenuation at lower temperatures to achieve noise-free signals. 
The first optical interconnect with a superconducting qubit detected the average optical power emitted from the qubit - a destructive measurement that prevented further use of the qubit state~\cite{mirhosseini2020}.
%was capable to optically detect the averaged qubit state, but this scheme was destructive for the qubit itself and hinders further manipulation of the qubit state . 
%An optical qubit state detection with l
Low back-action qubit readout has also recently been shown with a mechanically mediated electro-optical interconnect
%with a mechanical mediator has shown single-shot fidelity
~\cite{delaney2022} in a scheme comparable to Fig.~\ref{fig:1}b, but this relatively low bandwidth method necessitates additional microwave pumps with the associated heat load and isolation requirements.
%and . However, the necessity of a strong, noise-free microwave pump for the converter prevents a reduction in heat load and allows only an optical readout of signals from the qubit in contrast to the optical generation of microwave signals for qubit readout or control.
Ultra-high bandwidth readout of an electro-mechanical system has been demonstrated with a commercial electro-optic modulator operated at 4 K 
%has been used for an averaged optical 
but with limited efficiency and noise performance~\cite{youssefi2021}.
On the input side,
%a promising approach based on commercial high-bandwidth components is the use of 
%and providing high bandwidth uses 
high-speed %cryogenic 
photodetectors have been used to demodulate microwave control and readout signals
%for single shot qubit readout or qubit control 
~\cite{lecocq2021}.
This is a promising approach for multiplexed control but necessarily dissipative and does not allow to convert the readout signals back to the optical domain. 
%\textcolor{clue}{Recent power-efficient demonstrations of qubit control with an integrated electro-optic transducer \ and optical readout on an integrated piezo-opto-mechanical device show further potential for both denser packaging and heat load reduction alike for the input and output path respectively} 
%One drawback is the active power dissipation accompanying optical (destructive) photodetectors. This puts constraints on the scalability as the dissipation happens at the coldest stage of the dilution refrigerator which features the lowest cooling power. Additionally, this work did not demonstrate the optical detection of microwave signals reflected from the qubit.\textcolor{blue}{The previous sentence doesn't make sense to me. Maybe better: Additionally, the optically generated microwave signals could not be converted back to optical signals after interacting with the qubit.} In conclusion, a scheme that sends and detects optical signals for qubit readout, i.e. an all-optical readout, which could drastically reduce the number of refrigerator components, remains an open challenge. 
%\textcolor{blue}{I am tempted to use here generation and detection signals but we need to be careful a bit with the word generation because technically the signal is generated by an RF sourced and optics is only modulated} 

%\vspace{0.25cm}
%\noindent
%\textbf{Comparing conventional microwave with photonic readout setups}\\
In this paper, we 
% use a single electro-optic link to simultaneously send and detect RF modulated laser light enabling an 
demonstrate all-optical single-shot readout of a superconducting qubit, i.e. we replace both the input and output signal path by one optical fiber each, as shown in~Fig.~\ref{fig:1}c. 
Using a single electro-optic interconnect, i.e. a triply resonant whispering gallery mode single-sideband transducer~\cite{Rueda2016,hease2020,Fan2018,Sahu2023},
we simultaneously modulate and demodulate the optical carrier at millikelvin temperatures. This allows for a novel circulator-free readout that is used for time-domain characterization of a superconducting transmon qubit enclosed in a 3D superconducting cavity (cQED system)~\cite{paik2011}. 
%- the circuit QED (cQED) system. 
The latter is directly connected to the EO transceiver via a short coaxial cable without the need of any other cryogenic microwave components.
The ability to perform both microwave and optical measurements allows to make a quantitative comparison of the assignment fidelity of different readout types. 
It also enables sensitive Josephson parametric amplifier measurements in the presence of the readout laser. 
We employ that to carefully quantify the potential radiation 
%impact 
%(QND-ness) as well as the 
and average thermal impact on the mode occupancy and coherence of a superconducting processor that is operated with laser light.%\LQ{QND undefined}

\begin{figure*}[t]
	\includegraphics[scale=0.92]{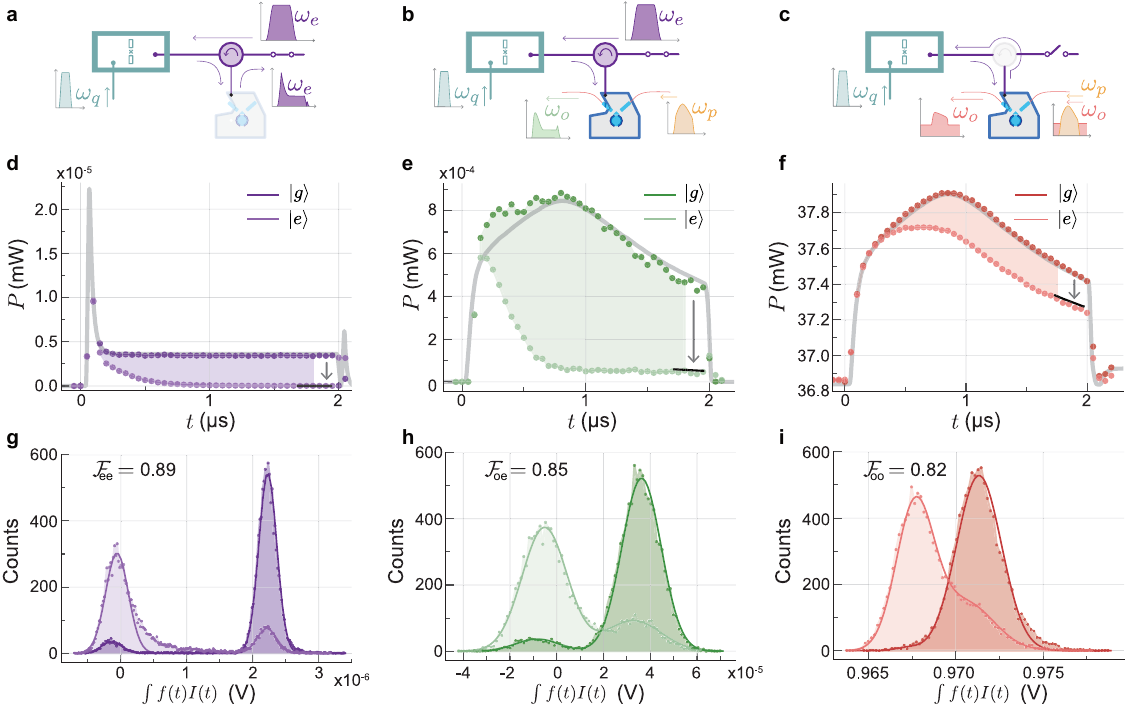}
	\caption{\textbf{Conventional and optical          %\LQ{optical, consistent with title} 
 single-shot readout of a superconducting qubit.} 
	   \textbf{a,b,c}, Sketches of the different readout schemes involving a microwave cavity dispersively coupled to a transmon qubit (cQED system in jade) and the electro-optical transceiver, consisting of a second microwave cavity (blue/gray) coupled to an optical whispering gallery mode resonator (light blue). 
        %Both microwave cavities are tuned on resonance with the free spectral range of the optical resonator ($\omega_{e} = \omega_{c} = \omega_{\mathrm{FSR}} \approx \, 2\pi \, 8.8 \, \mathrm{GHz}$). 
        The qubit state is prepared via a separate port at $\omega_{q}$. The electro-optic transceiver is operated with an optical pump pulse at $\omega_{p}$ to parametrically enhance the interconversion of microwave $\omega_{e}$ and optical $\omega_{o}$ signals. 
        \textbf{a}, Conventional microwave readout: a microwave pulse probes the cQED system and is detected via microwave heterodyne detection.
        %\LQ{MW heterodyning seems to be quite casual here. Nature seems to prefer detailed caption, i.e. you don't have repeat the same description in the main text}
        \textbf{b}, Optical detection of a microwave readout tone: the microwave pulse reflected from the cQED system is upconverted to the optical domain and detected with optical heterodyne detection. 
        \textbf{c}, All-optical readout: a modulated optical carrier is converted to the microwave domain
        %by a pump pulse $\omega_{p}$, interacts with 
        to probe the cQED system. Its reflection is simultaneously converted back to the optical domain and detected with an optical heterodyne setup. %\LQ{confusing sentence.}
        \textbf{d, e, f}, Averaged time traces of the correspondingly measured heterodyne signal powers for the qubit prepared either in the $|g\rangle$ or $|e\rangle$ state based on 15,000 independent measurements. 
        %Detected microwave (\textbf{d}) and optical (\textbf{e}) heterodyne signal from a microwave tone reflected from the cQED system. (\textbf{f}) Detected optical amplitude from a reflected RF modulated optical input signal after demodulation, microwave cQED probing and modulation. 
        The shaded area highlights the difference between both qubit states, which serves as an appropriate weighting function $f(t)$ for the time trace integration. 
        \textbf{g, h, i}, 
        %\LQ{For fidelity, maybe stick to the old convention $F_{oe}$, i.e. mw in and optics out? The letters are displaced in a weird way...}
        Corresponding histograms of 15,000 single-shots obtained by integrating the weighted in-phase quadrature $f(t) I(t)$
        %. weighted by $f(t)$ for a microwave signal and microwave detection readout (\textbf{g}), microwave-optical readout (\textbf{h}) and optical-optical qubit state detection (\textbf{i}) 
        with the corresponding state assignment fidelities $\mathcal{F}_{ij}$.    
	%\LQ{Remove grids in the plots? Panel numbers missing. Larger a,b, and c?} 
 }
      %  \textcolor{blue}{Make qubit pulse the same color as the cQED system, give the framge of the EO system also a color so I can refer to in the caption, CQED is cyan, The circulator in the last image should be made even more transparent, similar to the EO device in the first sketch, in general I find it a bit confusing the way components are fainted now. And we should put the microwave connector for the EO device on top so that the MW line doesnt cross the optical path. Its just a sketch, it doesnt need to be precise, the microwave line going around the circulator in the last page should have a larger radius, same distance to circulator like for straight line afterwards, put the qubit cavity connector a bit closer to the qubit but leave the qubit pump shape there}
  %\captionsetup{justification=justified}
  \label{fig:2}
\end{figure*}

%\section{Results}
%\LQ{Introduce the modular setup with key parameters first, and then discuss about the three configurations and respective measurements.\\
%Missing error bars for key numbers.}

%%%%%%%%%%%%%%%%%%%%%%%%
%%%%%%%%%%%%%%%%%%%%%%%%
\vspace{0.25cm}
\noindent
\textbf{Dynamics and single-shot fidelity of conventional and optical 
%\LQ{optical, consistent with title}
readout methods}\\
%%%%%%%%%%%%%%%%%%%%%%%%
%%%%%%%%%%%%%%%%%%%%%%%%
We start with a comparison of the three different readout methods schematically depicted in Fig.~\ref{fig:1} (see SI for detailed experimental setup):
%In order to compare the performance of these readout schemes, we perform identical measurements in three different configurations:
\begin{enumerate*}[label=(\roman*)]
  \item \emph{all-microwave readout} (Fig.~\ref{fig:2}a) with a microwave tone sent through coaxial cables to the cQED system and detected with a standard microwave heterodyne setup,
  \item \emph{microwave-optic readout} with optical detection of the same microwave signal as in (i) from the cQED component after using it for the modulation of laser light via the EO transceiver (Fig.~\ref{fig:2}b), and
  \item \emph{all-optical readout}: We send modulated light to the EO transceiver. The demodulated microwave pulse enters the cQED system and its reflection is converted back into the optical domain using the same EO transceiver 
  %again modulates the optical carrier 
  before being analyzed with an optical heterodyne detector at room temperature (Fig.~\ref{fig:2}c).
\end{enumerate*}
  All three schemes can be realized without setup changes except for the state of a cryogenic RF switch, as shown in Fig.~\ref{fig:2}a-c. While the first two methods can be performed simultaneously, opening the RF switch prevents the optically demodulated microwave signal in the all-optical readout from entering the microwave output line, which also effectively removes the circulator.
  %, cf.~Fig.~\ref{fig:2}c.
%\textcolor{blue}{GA Repetition of text to Fig.1 In principle I would like to point out the general setup reduction in Figure 1 with filters, amplifiers etc and in Figure 2a,b,c I would like to stress more how we did the measurements, i.e. no setup changes except for MW switch, what input, output signal were used etc. but it's hard to avoid repetitions}

%To minimize information loss the three components of the readout, i.e. the cavity resonance of the cQED system $\omega_{c}$, transducer cavity $\omega_{e}$ and the free-spectral range of the optical resonator of the transducer ($\omega_{\mathrm{FSR}}$), are tuned on resonance. This determines also the frequency of the microwave readout signal sent to the cQED system for the first two readout schemes as well as the modulation frequency of the optical signal used for the all-optical readout.   

The operation frequency of the photonic link is determined by its optical free-spectral range $\omega_{\mathrm{FSR}}/(2\pi) = 8.8065~\mathrm{GHz}$ set by the diameter of the lithium niobate resonator. To achieve a triply-resonant configuration that maximizes the transduction efficiency, we tune the electro-optic microwave cavity in resonance $\omega_{\mathrm{e}}=\omega_{\mathrm{FSR}}$ \cite{Rueda2016}. Similarly, to maximize the dispersive qubit readout efficiency \cite{Wallraff2005} we also tune the cQED cavity to the same frequency $\omega_{\mathrm{c}}=\omega_{\mathrm{FSR}}$. Both are implemented with a piezoelectric actuator.
%\LQ{piezo or piezoelectric}
%The cQED cavity and the microwave cavity of the electro-optic system are tuned on resonance ($\omega_{\mathrm{FSR}}=\omega_{\mathrm{e}}=\omega_{\mathrm{c}}\approx2\pi~8.8~\mathrm{GHz} $) with a piezo-mechanical actuator. 

The transmon qubit with anharmonicity $\nu/(2\pi)=201$~MHz is alternately prepared in its first excited state $|e\rangle$ or thermalized in its ground state $|g\rangle$ by selectively applying a flat-top-Gaussian microwave pulse of duration 104 ns
%of duration $\tau = \pi/\Omega_{R}$ 
at the qubit transition frequency $\omega_\mathrm{q}/(2\pi)=6.625~\mathrm{GHz}$ via a dedicated drive line, as shown in Fig.~\ref{fig:2}a-c. The readout tone, on the other hand, is either applied via filtered and attenuated input coaxial lines (Fig.~\ref{fig:2}a-b) or directly generated by the electro-optic transceiver (Fig.~\ref{fig:2}c) via resonantly enhanced optical down-conversion~\cite{sahu2022}. The readout amplitude corresponding to approximately $\sqrt{n_\mathrm{meas}}=122~\mathrm{photons}^{1/2}$ in the cavity is chosen to optimally benefit from the Jaynes-Cummings nonlinearity of the qubit-cavity system~\cite{Bishop2010,Boissonneault2010} that maps the qubit-state-dependent dispersive frequency shift of the resonator $\chi/(2\pi)=6.6$~MHz
%\LQ{$\chi$ in SI} 
into a large readout amplitude difference at the bare cQED cavity frequency $\omega_e$, see SI for details. The latter allows for %a (destructive) 
single-shot readout of the qubit state without a quantum-limited amplifier~\cite{Reed2010a}.

Figure~\ref{fig:2}d shows the ensemble averaged reflection amplitude from heterodyne detection in power units for the all-microwave readout. 
%scattering parameter $\langle|S_{11}|^2\rangle$ reflected microwave signal power. 
The measured dynamics with the qubit initialized in its ground state is in excellent agreement with the input-output relations taking into account the reflection from the transducer microwave cavity 
%reflection 
(gray line, see SI) as part of the signal output path.
When the qubit is prepared in the excited state, the nonlinear dynamics switches the cQED cavity reliably to its bare frequency for the same readout power, which effectively results in a critically coupled configuration and a vanishing reflection. While the dynamics are out of reach to be modeled given the high photon numbers, we adopt a simple cascaded cavity model between the cQED cavity and the EO microwave cavity~\cite{C.W.Gardiner1992}.
This accurately predicts the steady-state result after times $> 1.0~\mu$s without free parameters (black line highlighted with an arrow, cf.~also SI) and consequently the readout contrast between both states.
We then use these averaged measurements to find the optimal quadrature rotation and the integration weights (shaded region in Fig.~\ref{fig:2}d) that maximize the distinguishability for the single-shot-readout. 

%Integrating the individual quadrature signal encodes information about the qubit state. This can be optimized by an according quadrature rotation to maximize the distinguishability and the use of integration weights given by the difference of the averaged time trace between the two qubit states (shaded region in Fig.~\ref{fig:2}d).
%Conscipciously, this is only reached after $\apporx~1~\mu s$ and thereby increases the required readout length. In a low power readout dispersive readout on the same sample, we reach the steady state in less than half of the time and thereby could use short optical pulses. 

%We show the corresponding histograms from $1.5 \cdot 10^4$ measurements of each qubit state in  Fig.~\ref{fig:2}g by integrating the in-phase quadrature after maximizing the distinguishability by an according quadrature rotation. Each single-shot trace is weighted by the difference of the averaged time trace between the two qubit states and integrated over the first 1800~ns of the pulse (shaded region in Fig.~\ref{fig:2}d). 

We show the corresponding single-shot histograms from $1.5 \times 10^4$ independent measurements for each qubit state in Fig.~\ref{fig:2}g with double-Gaussian fits~\cite{Walter2017}. The maximum state assignment fidelity of 
$\mathcal{F}_{ee}=1-\left(P(e|g)+
P(g|e)\right)/2=0.89 \pm 0.01 $ is reached after an integration time of 1.8~$\mu$s, with $P(x|y)$ being the probability to measure the qubit in state $\left|x\right>$ after preparation of state $\left|y\right>$.
%\LQ{typo in the equation}to correctly assign the qubit state. 
The clear separation between the two distributions indicates a negligible overlap error ($\epsilon_{ol,e}<10^{-10}$).
The ground state error ($\epsilon_{g,e}\approx~7~\%$) originates partly from thermal excitation ($1.5~\%$ as quantified below),
while the rest is attributed to transitions induced by the comparably long high power readout pulse~\cite{Sank2016}.
The excited state readout results in an error of $\epsilon_{e,e}\approx~16~\%$, which suffers additionally from qubit decay during the measurement, shown as an asymmetric tail in the excited state Gaussian towards the ground state distribution. 
%Again, a fraction of the total excited state error $\epsilon_{e,e}\approx~16~\%$ can neither be explained by spontaneous emission nor thermal excitation and has to be attributed to the readout tone.

For a direct comparison, we simultaneously also read out a small part of the reflected microwave readout tone optically, as shown in Fig.~\ref{fig:2}b. 
After resonantly enhanced microwave to optical conversion~\cite{sahu2022}, in which about 3\% of the intra-cavity microwave photons are converted, we perform optical heterodyne detection, which yields the averaged time traces shown in Fig.~\ref{fig:2}e. 
In comparison to the microwave readout, we find slower dynamics and an additional decay of the optical readout signal at $\omega_o/(2\pi)=193.4$~THz. 
This is due to 
%readout via the photonic link shows a distortion of the original microwave modulation signal in the averaged optical power at frequency $\omega_o=193.4~THz$ by a) 
the limited conversion bandwidth of $\approx$~10~MHz 
%of the electro-optic link and b) 
and the particular shape of the $\approx 140~$mW parametric optical pump pulse at frequency $\omega_p = \omega_o - \omega_{\mathrm{FSR}}$ in very good agreement with theory (gray line) and the steady-state prediction (black line). %\LQ{This sentence is unclear. You mean g and e?}
%to the the photonic link enabling the efficient electro-optic modulation (Fig.~\ref{fig:2}e). 
As a result, the separation between the single-shot state distributions decreases (Fig.~\ref{fig:2}h), resulting in a larger overlap error of $\epsilon_{ol,eo}=2\%$ and a slightly reduced microwave-optical state assignment fidelity of %$\mathcal{F}_{oe}=0.849 \pm 0.009$.
$\mathcal{F}_{oe}=0.85 \pm 0.01$. 
%\LQ{FOE}.
% (Fig.~\ref{fig:2}h). 

Finally, also in case of the all-optical readout the optically demodulated microwave tone (corresponding to $\sqrt{n_\mathrm{meas}}=116~\mathrm{photons}^{1/2}$ in the cQED cavity)
%can detect the qubit state and 
results in well-distinguished state dependent trajectories as shown in Fig.~\ref{fig:2}f. The large optical background signal is due to the 
%the absence of an optical circulator leads to the detection of the 
cumulative reflection of the optical input e.g. at the coupling prism. 
%and thereby a large measurement background Fig.~\ref{fig:2}f. 
%\LQ{Maybe start with we show optical blabla in Fig.2f, which displays large background due to ... I think the optical circulator is not relevant here.}
%mainly from the coupling prism. 
The bandwidth of the EO transceiver now also slows down the dynamics of the build-up of microwave readout photons. 
%Nevertheless, when the optical pump pulse is turned on, 
Taking also into account the effect of electro-optically induced transparency 
%a small cavity linewidth change due to coherent electro-optical backaction
~\cite{Qiu2023},
%\LQ{EOIT}
which raises both signal levels during the optical pulse, the data is 
%increased background signal due to  and 
in excellent agreement with theory (gray and black lines).
%\LQ{Is there even DBA, since the optical modes are so symmetric.}
The moderate reduction of fidelity $\mathcal{F}_{oo}=0.82 \pm 0.01$ can be fully attributed to the larger overlap error between the state distributions shown in Fig.~\ref{fig:2}i. This result proves the feasibility of an isolator-free qubit readout without cryogenic microwave components.

\begin{figure}[t!]
\includegraphics[scale=0.9]{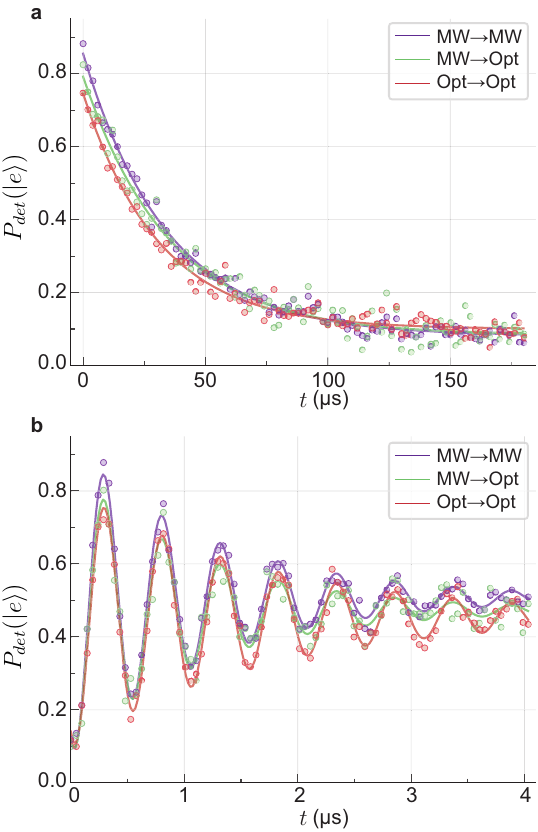}
\caption{\textbf{Qubit coherence for different readout methods.} 
        %\LQ{$P_{det}$ was used in Fig.2 for detected power. also add Delta to tau}
		\textbf{a}, Measured excited state detection probability $P_\textrm{det}(|e\rangle)$ 
  %of the qubit state 
  after a $\pi$ pulse for varying measurement delays $t$
  %\Delta \tau$ 
  using the three different readout methods shown in Fig.~\ref{fig:2}.
%  for conventional microwave detection (red), optical detection of a microwave readout tone (green) and a full optical readout with an optical signal and an optical heterodyne detection (blue) 
\textbf{b}, Measured Ramsey oscillations using two $\pi/2$ pulses separated by a variable delay $t$ 
%$\Delta \tau$ 
and detuned by $\approx2$~MHz from the qubit transition for the 
%coherence of the qubit measured via a T2 measurement 
 three readout methods.
		}\label{fig:3}
\end{figure}

\vspace{0.25cm}
\noindent
\textbf{Time-dependent qubit measurements}\\
%%%%%%%%%%%%%%%%%%%%%%%%
%%%%%%%%%%%%%%%%%%%%%%%%
We use all three readout methods to extract the longitudinal $T_1$ and transverse relaxation time $T_2^*$ of the superconducting qubit, %\LQ{i.e. T1 and T2,} 
based on 2,000 preparations with a 10~Hz repetition rate. 
%from an average over 2000 measurements with all three readout schemes. 
%The extracted T2s from fits are 1.27+-0.04 (MW), 1.27 +- 0.08 (MW2Opt), 1.73 +- 0.07 (Opt2Opt). The extracted T1 from fits are: 35.4+-1.6 (MW), 34.5+-0.08 MW2Opt and 32+-1 for OptOpt. So OptOpt has slightly worse T1 but better T2. 
Figure~\ref{fig:3}a shows the energy relaxation after a 
%A sweep of the time delay between a Gaussian 
$\pi$-pulse 
%of length $104~ns$ to prepare it in the excited state and the readout measurement yields a mean longitudinal decay 
yielding a consistent 
$T_1$ of $31.1$ to $35.4~\mu$s
%(33.7 \pm 1.7 )
for the three readout types.
%\LQ{maybe don't quote the number here. or $\sim 33.7$ without error bar. In addition, any reason for brackets?} 
The observed differences are in line with expected $T_1$ variability over the course of days. The individual signal-to-noise ratio and resulting confidence level is very similar for all three measurements and the slightly reduced contrast is expected due to the previously extracted $\mathcal{F}_{ij}$.
%single-shot state assignment fidelities.     
%(Fig.\ref{fig:3}a). The varying contrast between the measured qubit states for different readout types results from the slight differences in the readout fidelity calculated from single shot analysis. 

Similar conclusions can be drawn from the measured exponential decay of the Ramsey oscillations shown in 
%as a function of the delay between two $\pi/2$ pulses 
Fig.~\ref{fig:3}b. 
The fitted mean transverse decays for all three measurements, $T_2^*=1.16-1.73~\mu$s, 
%(1.4 \pm 0.3 )
are comparable with the all-optical readout yielding the longest coherence. 
%with no sign of a radiative impact. 
%\LQ{Same here.}
We attribute the comparably low $T_2^*$ of this particular device to fabrication and design-related issues,
as the theoretical limit due to energy decay 
%($e^{-t/2T_1}$) 
and dephasing from thermal cavity shot noise~\cite{Clerk2007} is estimated to be $T_2^* \approx 20.0~\mu$s and 
a Hahn-echo measurement of $T_{2,\mathrm{echo}} = (1.40 \pm 0.03 )~\mu$s excludes a low frequency, e.g. mechanical, noise origin.
%This is supported by a strong variation of the transversal relaxation times in the transmon fabrication for this work. 
%Although the origin of the low phase coherence needs further investigation for future experiments requiring improved qubit lifetimes, 
Moreover, we observe the same coherence times when the readout laser is turned off, or 
%alternatively 
when the optical pulse is applied during the qubit state preparation, as discussed below.
Our measurements, therefore, clearly demonstrate the integrity of superconducting qubit coherence using a photonic readout.

%in the presence of the electro-optic link in operation.

%JF: 
%- single shot analysis (also for the y axis?)?... then its not averages
%- use a different variable for the y axis
%- state the pi pulse type and length
%- maybe (maybe not) state the fidelities we know of (state prep. readout. etc.) - ist this the reason for the reduce contrast? then comment on how small the difference is and why.
%- what is the result? state all timescales with error bars
%- comment on what it means. are the values large or small and why? do you see a difference in confidence or mean of the different measurement types?
%- if no, what does that mean?

%%%%%%%%%%%%%%%%%%%%%%%%
%%%%%%%%%%%%%%%%%%%%%%%%
\vspace{0.25cm}
\noindent
\textbf{Impact of optical absorption heating 
quantified with quantum-non-demolition measurements
} 
%on QND-ness, qubit time-scales and mode occupancies}
\\
%%%%%%%%%%%%%%%%%%%%%%%%
%%%%%%%%%%%%%%%%%%%%%%%%
While the previous measurements have shown that reliable qubit characterization is feasible 
with a strong optical readout pulse, a more sensitive method is required to fully quantify the potential radiative~\cite{Barends2011,Houzet2019} and thermal~\cite{hease2020} impact of high energy pump photons. 
%pump with peak powers up to 180 mW at the input of the optical resonator, 
%the exact impact on the qubit coherence needs to be investigated, especially considering the shown degradation of qubit properties due to quasiparticle generation of high energy photons~\cite{Barends2011,Houzet2019}. Besides stray light, photons can also corrupt the qubit indirectly by thermal noise as a result of heating due to optical absorption. While stray light is only present during the optical pulse, absorption heating can persist over long time scales of the order of minutes due to the large specific heat of our mm-size electro-optic system~\cite{hease2020}. 
In the following, we use a near-quantum-limited non-degenerate Josephson parametric amplifier~\cite{Winkel2020} to perform a low power, dispersive, and non-destructive qubit readout to quantify such effects.

\begin{figure}[!]
\includegraphics[width=0.95\linewidth]{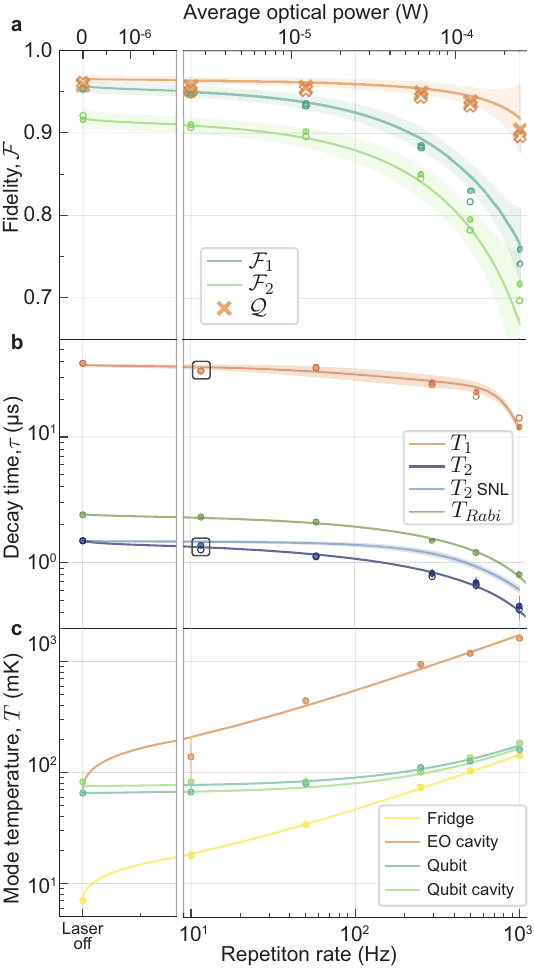}
\caption{\textbf{Impact of the optical pump.} 	
  \textbf{a}, Measured state assignment fidelities of two consecutive JPA-assisted microwave measurements (dots) and corresponding QND metric (crosses) obtained in the presence of a $2~\mu$s long optical pump pulse of $\approx 0.14~$W applied during the first readout as a function of repetition rate and calculated average optical power (top axis) together with theory (lines and $3\sigma$ confidence bands). Open symbols (mostly overlapping) denote measurements where the optical pump was applied during state preparation.  %for increasing repetition rates of the optical pump (bottom axis) and average optical pump powers (top axis). 
  %Theory (lines)
  %predictions comprise 
  %takes into account the measured thermal excitation and spontaneous emission as well as 
  %for the measured temperatures shown in panel c. Measurement-induced mixing as the only free fit parameter
  %, e.g. due to radiation, 
  %is $<1\%$ measurement-induced mixing.
  \textbf{b}, Measured qubit coherence times when the optical pulse is synchronized with each state preparation and readout (open circles) and for a free-running measurement sequence (dots) vs. optical pulse repetition rate. Squares indicate the mean of the optical readout results in Fig.~\ref{fig:3}.
  %with a $3\sigma$ statistical error bar.
  %vs. 
  %optical pulse 
  %repetition rate. 
  %The qubit coherence significantly decreases for larger rates along the longitudinal and transverse direction. 
  The decrease in $T_1$ is accurately modelled with theory (red line and $3\sigma$ confidence band) but the shot noise dephasing limit (blue line) does not capture the measured $T_2$ dependence, 
  %the measured thermal occupancy shown in panel c, the expected quasiparticle distribution and Purcell decay (red line). Shotnoise dephasing cannot explain the measured $T_2^*$ (light  blue line), 
  which follows a 
  %. Other lines are powerlaws and serve as a guide to the eye. While the decrease in T1 is proportional to $\propto \, -x^{0.5}$, T2 can be described by 
  $\propto \, -x^{0.37}$ power law (dark blue line) similar to the measured Rabi decay (green). 
  \textbf{c}, Measured temperature of the mixing chamber (yellow dots) and the different microwave modes (dots) 
  %with $2\sigma$ error bars) 
  together with power law fits as a guide to the eye.
  %as a function of optical pulse repetition rate. 
  %Without optical pump 
  %%($\overline{P}_{opt}$ \,= \,0) the 
  %all modes
  %%components of the readout line 
  %thermalize to around 70~mK. 
  %%at the refrigerator base temperature of 7~mK. 
  %%%%%%%%%%%%%%%%%%
  %The EO-transducer cavity acts as heat source that %raises the refrigerator temperature from 7~mK
  %%- both in direct thermal contact - 
  %following the same power law $\propto \, x^{0.54}$ %(yellow and orange lines). The qubit and qubit %cavity temperature 
  %%remain relatively constant and only 
  %also start to rise once the 
  %%increasing 
  %refrigerator temperature approaches their %thermalization temperature of about 70~mK.  
		%\LQ{Merge the upper and lower panels (same x axis). Use consistent font and colors.}
  }\label{fig:4}
\end{figure}

First, we measure the alternately prepared qubit states two times back to back, the first one in the presence of the previously used optical pump pulse and for comparison also when the laser is off. Figure~\ref{fig:4}a shows the extracted assignment fidelity $\mathcal{F}$ of the first (second) measurement in cyan (green) for increasing optical pulse repetition rates. The observed dependence on the resulting applied average optical power (top axis) is in excellent agreement with theory (lines and $3\sigma$ confidence bands) for 
%confirmed by a convincing agreement of the measured fidelities with values expected from 
spontaneous emission scaling with $1-e^{-t/T_1}$ and the independently measured thermal excitation of the qubit, cf.~Fig.~\ref{fig:4}c. The remaining discrepancy is fitted to be $\leq 1\%$ and attributed to either measurement (or optical radiation) induced transitions or state preparation errors.
%The remaining discrepancy owing to measurement-induced transitions or state preparation errors is 

The quantum-non-demolition (QND) metric is defined as the fraction of measurements, where two consecutive measurements yield the same qubit state~\cite{Touzard2019}, i.e. ~$\mathcal{Q}=\left(P(g_\mathrm{2}|g_\mathrm{1})+P(e_\mathrm{2}|e_\mathrm{1})\right)/2$.
%i.e.~$\mathcal{F}_\mathrm{QND}=\left(P(g_\mathrm{meas2}|g_\mathrm{meas1})+P(e_\mathrm{meas2}|e_\mathrm{meas1})\right)/2$.
%Importantly, $\mathcal{F}_\mathrm{QND}$ (orange) is comparable to the state assignment fidelity of the second readout, which implies a minimal (if any) direct impact of the optical pulse on the qubit. This interpretation is backed by additional measurements where the same optical pulse is applied %not only during the first microwave readout but also 
%during the qubit state preparation (open symbols in Fig.~\ref{fig:4}), which mostly overlap.
 %This is further 
%It is ultimately limited by transitions during the readout.
Importantly, $\mathcal{Q}$ (orange) is comparable for moderate repetition rates and a dark measurement without laser light, which implies a minimal (if any) direct impact of the optical pulse on the qubit. This interpretation is backed by additional measurements where the same optical pulse is applied during the qubit state preparation (open symbols in Fig.~\ref{fig:4}), which mostly overlap. The theoretical prediction is based on spontaneous emission.
%We want to stress that the cQED system is directly connected to the photonic link via microwave cables and no additional shielding measures were taken apart from a conventional radiation shield enclosing the entire mixing chamber plate. %\textcolor{blue}{Which shield again?} shield around the qubit-cavity system.

%The coherence measurements in Fig.\ref{fig:3} show - besides the equivalence of all three readout methods to extract the qubit coherence times - the impact of both noise mechanisms, stray light and absorption heating, because the readout is performed simultaneously with the optical pump ("instantaneous heating"). In contrast, Fig.\ref{fig:4}a shows the qubit coherence from consecutive microwave measurements not synchronized with the optical pump. Thus, the extracted coherence time represents the impact of the mean temperature between two optical pulses ("averaged heating").First of all, the results are in complete agreement with Fig.\ref{fig:3} shown as black markers. This suggest that the qubit is not exposed to stray light from the electro-optic link and instantaneous heating is negligible. This fact is remarkable because we refrained from taking any measures to shield the qubit from the incoming optical signal of the all-optical readout. 

Figure~\ref{fig:4}b shows the identically measured qubit coherence times as a function of optical pulse repetition rate (open circles) together with free-running measurements where the optical pulse is not synchronized with the microwave measurements ($\approx$ 5 kHz repetition rate,
%that we repeat every 200~$\mu$s 
solid circles). The latter method is faster but would not reveal an instantaneous radiation-based impact, e.g. via the generation of quasiparticles. The very close agreement between the two types of measurements (most symbols overlap), the small difference between low repetition rates and laser-off measurements, and the very good agreement with the coherence times obtained with the optical readout of Fig.~\ref{fig:3} (black squares) indicates the absence of such radiative effects.  
%In the following step, we abolish the condition that the qubit is measured simultaneously with the optical pump and perform qubit measurements every 200$\mu s$ asynchronous with the optical pulse for varying optical repetition rates. Hence, the qubit is only susceptible to pump effects with large time constants on the order of the time between two optical pulses ($ \geq~1~ms$) and we measure the qubit's longitudinal and transverse decay $T_1$ and $T_2^*$. Similar to Fig.\ref{fig:4}a, moderate repetition rates are comparable to a cold environment when the laser is switched off. 
%As a comparison, the coherence times from Fig.\ref{fig:3} using optical readouts are indicated by black squares. The agreement highlights again 1) the consistency between measurements with the presented optical readout schemes and a quantum-limited amplifier assisted microwave readout, and 2) 

%The effect of the optical pump pulse is purely given by the steady state temperature as a consequence of slow heating mechanisms.

%When the repetition rate of the optical pulse is increased further, the average optical pump power in the electro-optic link rises and both the longitudinal and transverse qubit coherence decrease. This reveals a major drawback of this first proof-of-principle experiment: The optical pump power required for this particular electro-optical sample with an optical resonator of an only moderate quality factor $Q_{int}~=~5 \cdot 10^6$ allows only a low repetition rate before the qubit coherence degrades. 

The measured reduction in $T_1$ as a function of applied average optical power is in excellent agreement with optical absorption heating in the EO transducer. 
The theory curve in Fig.~\ref{fig:4}b (red line and $3\sigma$ confidence band) takes into account 
thermal radiation \cite{Corcoles2011} and thermal equilibrium quasiparticles with a superconducting gap of $\Delta = 205~ \mu$eV using the independently measured qubit temperature shown in Fig.~\ref{fig:4}c, as well as a typical nonequilibrium quasiparticle density of $1 \times 10^{-7}$~\cite{Serniak2018}, and an increase in the Purcell rate due to a somewhat broadened cQED cavity linewidth at higher repetition rates and temperatures. The relative dependence of the transverse decay $T_2^*$, however, cannot be explained by an increase in shotnoise dephasing from the thermally populated cQED cavity~\cite{Clerk2007} (light blue line), similar to its absolute value (cf. discussion Fig.~\ref{fig:3}b). $T_\mathrm{2,echo}$ and $T_2^*$ show again no measureable difference. Quasiparticles are also not believed to have a dominating effect on dephasing in transmon qubits~\cite{Catelani2012,Zanker2015,Riste2013}.

Finally, we investigate the average temperature distribution of the different components, which is used for the theory in Fig.~\ref{fig:4}a and b. Fig.~\ref{fig:4}c shows the measured base plate temperature from a calibrated rutenium oxide sensor, as well as the mode temperature of the superconducting qubit as obtained from thermally excited $|e\rangle \leftrightarrow |f\rangle$ Rabi oscillations \cite{Jin2015}. The temperature of the cQED cavity is extracted from populated Ramsey oscillations \cite{Dassonneville2021}, and the EO  microwave cavity temperature is calculated from the measured power spectral density at its output \cite{hease2020}. These measurements were performed free-running but with the same optical pulse applied to the transducer.

When the laser is off, all components thermalize to a temperature of $\approx$~75~mK, while the refrigerator reaches a base temperature of $\approx$~7~mK, see Fig.~\ref{fig:4}c. When the optical pump is on, it acts as a localized heat source that increases the EO microwave mode temperature (orange). The proportionality to the time-averaged applied optical power of $\propto \, \Bar{P}_\mathrm{opt}^{0.54}$ agrees with previous findings for continuous wave optical pumps \cite{hease2020}. 
The EO transceiver is in very good thermal contact to the refrigerator's base plate, which 
%. The temperature data suggests that the electro-optic device heats up the mixing chamber of the dilution unit 
heats up the refrigerator with the same power law (yellow).
%at a similar rate with respect to optical power, competing with the cooling power of the refrigerator. 
The resilience of the cQED system to radiation and heating at moderate repetition rates (cf.~Fig.~\ref{fig:4}a and b) is reflected again in the mode temperature of 
%its components, 
the qubit and the dispersively coupled cavity. Their temperature increases only slightly compared to the laser-off situation for moderate repetition rates. One reason for this behavior is the detuning between the transducer cavity mode and the cQED system by the Lamb shift $\chi_0/(2\pi)=26~$MHz except for the moment when the high power readout pulse is applied. Other reasons are the careful thermalization of all components and the large heat capacity and thermal contact area of the bulk EO transducer compared to integrated photonics approaches. %the microwave components in between. 
However, as the cQED system is thermally connected to the mixing chamber as well, its mode temperature rises as soon as the fridge temperature approaches the thermalization temperature of the cQED unit (cyan and light green). This behavior is consistent with the sharp decline in the qubit coherence and readout fidelity for higher repetition rates in Fig.~\ref{fig:4}a and b.

%%%%%%%%%%%%%%%%%%%%%%%%
%%%%%%%%%%%%%%%%%%%%%%%%
\vspace{0.25cm}
\noindent
\textbf{Conclusions and prospects}\\
%%%%%%%%%%%%%%%%%%%%%%%%
%%%%%%%%%%%%%%%%%%%%%%%%
%JF: Now about heat load:
One of the main motivation for this work is to simplify the cryogenic measurement setup by eliminating bulky and costly microwave components that are the source of a significant heat load \cite{Krinner2019}. %\LQ{Polish this sentence..}
%On the other hand, optical fibers as input and output paths eliminate the passive heat load of typical microwave cables from all temperature stages of a dilution refrigerator. 
In contrast, even the smallest cooling power at the mixing chamber plate can handle the passive heat load of millions of fibers \cite{lecocq2021} and their %the . In addition, the 
small cross-section %of optical fibers ($125 \mu m$) reduces 
mitigates the problem of space constraints raised by mm-sized coaxial cables. Nevertheless, the active heat load of this proof-of-principle all-optical readout limits the duty cycle and prevents a direct scaling-up to many readout-out lines.
%of this approach. 
%and we show that it also limits the rep rate. 
In the present case, due to our low optical coupling efficiency of $\eta_o$ = 0.22, a majority of the parametric pump power 
is absorbed at the mixing chamber, leading to the observed temperature increase associated degradation of the qubit coherence shown in 
%due to optical power dissipation competing with the mixing chamber's cooling power of $20\mu W$ (cf. 
Fig.~\ref{fig:4}. In the future, the optical coupling efficiency is therefore a critical parameter to improve, and optimized devices will also need to out-couple the majority of the reflected light to avoid absorption in the refrigerator. Similarly, the power efficiency is another critical parameter that can be improved dramatically with integrated photonic devices. One example is an electro-opto-mechanical device yielding a similar cooperativity for $10^{-9}$ times lower optical pump power \cite{arnold2020} albeit with lower bandwidth and noise performance.
%\LQ{I find it quite confusing to mention EOM here, in current way.
%Or:
%Similarly,... with intergrated photonics approaches. For example, on-chip electro-opto-mechanical devices yield a similar cooperativity for ... pump power.
%} %Sunch integrated photonics 

%A significant increase of optical coupling will therefore not only improve the detection efficiency (see above) but also drastically reduce the dissipated heat in the dilution unit. 

%JF: cut (coupling) losses and minimize pump power for C=1, i.e. maximize g and minimize kappa and gamma. there are a few orders of magnitude to win but integrated optics is the way to go down the road ...

%JF: mention hong tangs best work here for integrated. 

%A transition to integrated electro-optic links that have recently shown very power-efficient transduction~\cite{VanThiel2023, warner2023} will reduce the active heat load and increase the packaging density. Additionally, as long as the thermal noise remains small compared to the used signal, mounting these transducers on a higher temperature plate with more cooling power and better thermal conductivity is a promising option for a coherent measurement using electro-optical cryogenic interface. %For an averaged coherent measurement you could average the thermal noise away but for a single shot measurement, I think the coherent signal has to be bigger than the noise

%\section{Discussion and Outlook}
One of the limitations of the presented optical readout is the need for a comparably large number of readout photons $n_\mathrm{meas}$. %The performance of a detector is best quantified by its quantum efficiency. 
Scaling the histograms in Fig.~\ref{fig:2}g-i with the corresponding readout amplitude $\sqrt{n_\mathrm{meas}}$ 
%allows the extraction of the detection efficiency compared to an ideal, quantum limited amplifier. The scaled Gaussian variance of the histogram $\sigma_{det}^2$ and the variance of an ideal amplifier, $\sigma_{0}^2 = 0.5$, 
%determines the detection 
yields the quantum efficiency $\eta_\mathrm{det} = \sigma_{0}^2/\sigma_\mathrm{det}^2$ with the Gaussian variance of the measured histogram $\sigma_\mathrm{det}^2$ and the variance of an ideal phase insensitive amplifier $\sigma_{0}^2 = 0.5$ \cite{Hatridge2013}. 
%$\sigma_{0}^2 = 0.5$ is the variance of an ideal amplifier.
For the conventional microwave readout (without JPA) we extract
%has a detection efficency of 
$\eta_\mathrm{det,mw} \approx 1.3\times10^{-3}$.
%, referenced to the cQED system output. The dominant 
This is consistent with a comparably large amount of loss between the cQED system and the first amplifier (transmission of only $< 3 \%$) due to the extra circuit elements such as the EO converter with reflectivity $(1-2\eta_e)^2 = 0.09$, with the microwave coupling efficiency $\eta_e$. %\LQ{$\eta_e$ undefined in main text.}
%contributions stem from the thermal noise of the microwave HEMT amplifier, and the losses between the cQED system and the input of this amplifier $(< 3 \%)$.
%Comment: The transmission loss to the HEMT comes from EO cavity (only 0.13 reflectivity for eta = 0.35) then 2.5 dB cable losses to the MW switch and the rest are cable losses from MW switch to JPA and then to HEMT. The number is just calibrated from the HEMT noise from the  MW switch (16.65 port on reset) and the detection efficiency, i.e. 1/16.65*transmission to switch = det efficiency = 0.0013 ->eta from qubit cavity to MW switch = 0.021. But this is just a simple estimate, we don't need to state this number but can just say it is low due to EO cavity in the output path
%Besides cable losses, the small reflectivity of the on-resonant transducer cavity ($\eta_e = 0.35$) leads to such a low transmission to the HEMT amplifier. On the other hand, we 
For the two optical readouts on the other hand we find 
%corresponding optical readout efficiency without the optical counterpart of a HEMT amplifier of 
$\eta_\mathrm{det,opt} \approx 1.5 \times 10^{-4}$, which agrees with the moderate total electro-optic device conversion efficiency $\eta_{eo} = 0.3\%$ and optical losses. This is in the same ballpark as recent experiments with an electro-opto-mechanical system using a longer ($15~\mu s$) readout pulse \cite{delaney2022}. Importantly, 
%achieved in this work. 
%JF: Clarify these numbers with Georg. losses seem high (nadd >>10 here in this setup?), all opt. number not there. is the low power JPA readout close to the expected ideal limit?
%Although this optical detection efficiency is already comparable with the recently reported performance of an electro-opto-mechanical system \cite{delaney2022} using a longer $15~\mu s$ pulse for dispersive qubit readout, 
%Lehnert used a 15 us readout pulse, he used 381 photons (very small chi) and got a quantum efficiency of 8*1e-4, All these factors together made it possible that he can do single shot readout
%a superior electro-optic transducer such as in previous experiments 
even with just the original performance of this device \cite{sahu2022} (we observed optical Q degradation in repeated cooldowns) %would boost the efficiency by almost two orders of magnitude. This brings the detection efficiency in reach of readout tones with $\approx~100~$ photons, paving the way for an 
a %all-optical 
quantum-non-demolition single-shot readout without electronic amplifiers and readout times of $\approx~1~\mu$s would be within reach. With further realistic improvements of in/out coupling and transmission losses
%, as well as optical quality factor 
close to quantum limited detection efficiencies for photonic RF sensing \cite{SantamariaBotello2018} as well as high bandwidth and high fidelity qubit readout comparable to the state of the art \cite{Walter2017} will be possible.
%comparable to the state of the art 

%The small readout signals, however, would require improvements in our current optical output path (e.g. better filtering of the optical pump). Finally, our bandwidth of $\approx~10~\mathrm{MHz}$ can currently only compete with standard JPAs but not with the recent development of travelling wave paramteric amplifiers. This limitation prohibits a multiplexed alloptical qubit readout at the moment.

%JF: I'd rephrase a bit in the sense that we can do also better not only like before. and also not all above is relevant. i would mention that once we cut the losses between the two systems it behaves more like a resonator that can also use for state transfer and cite cleland.

%JF: we should also compare with state of the art in microwave readout (probably the Walter paper of the Wallraff group).

%JF: we should end on a high note. The positive message is that we show that qubits survive light for control and readout even though more sensitive meas. are needed (with more coherent qubits). This opens up new possibilities. maybe end with a modified version of this: Together with two other recent demonstrations on optical readout and control of a superconducting qubit [cite IST and Harvard paper], our work sets a comprehensive new direction for reducing the space and heatload constraints on superconducting quantum processors.

In summary, we have demonstrated a circulator-free %single-shot-readout of a 
superconducting qubit readout with an all-optical scheme that relies only on optical (de-)modulation and optical heterodyne detection. Such a platform offers a significantly simplified cryogenic setup where signal-conditioning is performed at room temperature %employing only a single electro-optic transducer at mK temperatures 
and optical fibers act as link to the cryogenic environment. 
%to the ambient environment. 
Somewhat surprisingly we found that the comparably high power optical pulse in the 100~mW range does not have a detrimental effect on the qubit coherence, despite the absence of shielding elements.
This result, when combined with recent integrated photonics demonstrations of more power-efficient and higher repetition rate optical control \cite{warner2023} and readout \cite{VanThiel2023} of planar superconducting qubits, provides a viable path towards all-integrated photonic operation of superconducting quantum processors.

\bibliography{Final_jf2v2}
\bibliographystyle{apsrev4-2}

%\vspace{0.25cm}
%\noindent
%\normalsize
%\textbf{Data availability}\\

%\vspace{0.25cm}
%\noindent
%\normalsize
%\textbf{Code availability}\\

\vspace{0.25cm}
\noindent
\normalsize
\textbf{Data and Code Availability}\\
The code and raw data used to produce the plots in this paper will be made available at the Zenodo open-access repository 
%under the link XYZ
at the time of publication.\\
% ~\cite{Knorzer2022}
%\LQ{arxivReference Format}

\vspace{0.25cm}
\noindent
\normalsize
\textbf{Acknowledgments}\\
We thank F. Hassani and M. Zemlicka for assistance with qubit design and high power readout respectively, and P. Winkel and I. Pop at KIT for providing the JPA. This work was supported by the European Research Council under grant agreement no.~758053 (ERC StG QUNNECT) and no.~101089099 (ERC CoG cQEO), the European Union's Horizon 2020 research and innovation program under grant agreement no.~899354 (FETopen SuperQuLAN) and the Austrian Science Fund (FWF) through BeyondC (grant no. F7105). L.Q.~acknowledges generous support from the ISTFELLOW programme and G.A.~is the recipient of a DOC fellowship of the Austrian Academy of Sciences at IST Austria.\\

\vspace{0.25cm}
\noindent
\normalsize
\textbf{Author Contributions}\\
 G.A. and T.W. performed the experiments together with R.S. and L.Q. G.A., R.S. and L.Q. developed the theory. G.A. and T.W. performed the data analysis. 
 %with help of R.S. and L.Q
 G.A. designed the qubit and developed the tunable microwave cavity, and L.K. fabricated the transmon qubit. The manuscript was written by G.A. and J.M.F. with assistance from all authors. J.M.F. supervised the project. \\

%L.Q. conceived the idea for the experiment. L.Q. and R.S. performed the experiments together with W.H. and G.A. L.Q. developed the theory and performed the data analysis. The manuscript was written by L.Q. with assistance from all authors. J.M.F. supervised the project. \\

\vspace{0.25cm}
\noindent
\normalsize
\textbf{Competing Interests} \\
The authors declare no competing interests.

%\LQ{SI is moved to a separate file.}

\end{document}

\title{
Supplementary information "All-optical single-shot readout of a superconducting qubit" 
%readout
%Photonic superconducting qubit readout
}
\author{Georg Arnold$^\star$}
\email{georg.arnold@ist.ac.at}
\author{Thomas Werner$^\star$}
\author{Rishabh Sahu}
\author{Lucky N. Kapoor}
\author{Liu Qiu}
%\author{Liu Qiu$^\star$}
%\author{Rishabh Sahu$
%\author{William Hease}
\author{Johannes M. Fink}
\email{jfink@ist.ac.at}
\affiliation{Institute of Science and Technology Austria, Am Campus 1, 3400 Klosterneuburg, Austria}

\date{\today}

\maketitle

\onecolumngrid
\tableofcontents
\addtocontents{toc}{~\hfill\textbf{Page}\par}
\newpage

%%%%%%%%%%%%%%%% SUPPLEMENTARY

\begin{table}[h!]
	\begin{tabular}{|c|p{12cm}|}
		\hline
		\multicolumn{2}{|c|}{\textbf{Introduced in Main Text}} \\ \hline
        $C$                                                                                   & cooperativity ( $ C= 4 g^2/\kappa_e\kappa_o$ )                                             \\ \hline
        $\chi_{g,e}$                                                                          & Dispersive qubit-state-dependent frequency shift                                                               \\ \hline
        $\chi_0$                                                                &  Lamb shift  
        \\ \hline
        $\Delta$                                                                          & Superconducting gap \\ \hline
         $\epsilon_{g,e},\epsilon_{e,e}$                                                                          & Ground state, excited state error                                                                \\ \hline
         $\epsilon_{ol,e},\epsilon_{ol,oe}$                                                                          & Overlap error for mw, mw$\rightarrow$opt              readout                                                  \\ \hline
        $\eta_{det},\eta_{det,mw},\eta_{det,opt}$                                           & Quantum efficiency for microwave, optical detection\\ \hline
		$\eta_{e}$                                                                          & External coupling efficiency transducer cavity                                                                  \\ \hline
		$\eta_{eo}$                                                                         & Total electro-optic conversion efficiency                                              \\ \hline
        $\sqrt{n_{meas}}$ / $n_{meas}$                                                                &  Readout amplitude / readout photon number  
        \\ \hline
		$\eta_{o}$                                                                               & External coupling efficiency optical resonator                                                     \\ \hline
      $f(t)$                                                                          & Weighting function for time-trace integration                                                               \\ \hline
        $\mathcal{F}_{ee},\mathcal{F}_{oe},\mathcal{F}_{oo}$                                                                          & Assignment fidelity for mw$\rightarrow$mw, mw$\rightarrow$opt, and opt$\rightarrow$opt     readout                                                          \\ \hline
        $|g\rangle, |e\rangle, |f\rangle$                                                                              & Qubit ground, excited, second excited state                                                                                  \\ \hline
        $I(t)$                                                                          & In-phase quadrature of the microwave output field                                                               \\ \hline
		mw $\rightarrow$ mw,mw $\rightarrow$ opt,opt $\rightarrow$ opt                                                                              & Sending and measuring microwaves, sending microwaves and measuring optics, sending and measuring optics                                                                                \\ \hline
         $\nu$                                                                              & Qubit level anharmonicity                                                                                  \\ \hline
          $\omega_{c}$                                                                              & Resonance frequency of the cQED system                                                                                  \\ \hline
        $\omega_{\mathrm{FSR}}$                                                                              & Optical free spectral range                                                                                  \\ \hline
        $\omega_{e}$                                                                          & Microwave resonance frequency of the electro-optic transducer                                                               \\ \hline
       $\omega_{o}$                                                                          & Optical signal frequency                                                               \\ \hline
        $\omega_{p}$                                                                          & Optical pump frequency                                                               \\ \hline
        $\omega_{q}$                                                                              & First qubit transition frequency                                                                                  \\ \hline
        
        $P(e|g)$ ($P(g|e)$)                                                                          & Probability for assigning $|e\rangle$($|g\rangle$), after preparing $|g\rangle$($|e\rangle$)                              \\ \hline
        
        $P(e_1|e_2)$ ($P(g_1|g_2)$)                                                                          & Probability to measure $|e\rangle$($|g\rangle$) in two successive measurements                                                   \\ \hline
        $P_{opt}$                                                                     &  Optical power           
        \\ \hline
        $\mathcal{Q}$                                                                          & Quantum non-demolition metric (QND-ness)                                                                \\ \hline
  		$\sigma^{2}_{det}$                                                                               & Scaled Gaussian variance                                                                      \\ \hline
		$\sigma^{2}_0$                                                                    &  Variance ideal amplifier\\ \hline
         $T$                                                                          & Mode temperature                                                                \\ \hline
         $T_{1}$                                                                          & Energy relaxation                                                                \\ \hline
         		$T_{2,echo}$                                                                  & Transverse echo relaxation                                         
  \\ \hline
         $T_{2}^*$                                                                          & Transverse decay                                                                \\ \hline
        \multicolumn{2}{|c|}{\textbf{Introduced in Supplementary information}} \\ \hline
        $\hat{a}_j$                                                            & Mode, $j\in(e,o,c,p,s,tm)$
		\\ \hline
        $\hat{a}_{j,\mathrm{in}}$                                           & Input field (noise) operator for the microwave and optical mode, $j\in(e,o,c)$\\ \hline
        $\hat{a}_{j,out}$                                                            & Readout field, $j\in(e,o,c)$
		\\ \hline
        $\delta_j$                                                            & Detuning, $j\in(o,s,tm)$
		\\ \hline
        $\eta_{j}$                                                            & External cavity coupling efficiency of individual mode, $j\in(e,o,c,p)$\\ \hline
        $\eta_{j,i}$                                                          & Coupling efficiency from i to j, $i,j\in(e,o,c)$
		\\ \hline
        $f_{bare},\omega_{bare}$                                                                   & Bare resonator frequency cQED, 2$\pi f_{bare}$ \\ \hline
        
        $g$                                                                                   & Photon enhanced electro-optical coupling rate ($g=\bar{a}_p g_0$)                 \\ \hline
        $g_0$                                                                                   & Electro-optic vacuum coupling rate ($g=\bar{a}_p g_0$)                 
		\\ \hline
        $J$                                                                   & Coupling rate between the optical Stokes mode and TM mode
        \\ \hline
        $\kappa_{j}$                                                            & Total loss rate of individual mode, $j\in(e,o,c,s,p)$
		\\ \hline
        $\kappa_{j,ex}$                                                            & External loss rate of individual mode, $j\in(e,o,c)$
		\\ \hline
        $m$                                                                   & Microwave azimuthal mode number, $m=1$
        \\ \hline 
        $P_{ee}$                                                                   & Detected power for mw$\rightarrow$mw
        \\ \hline
        $\hat{\sigma_z}$                                                            & Pauli operator
		\\ \hline
        $\tau$                                                                   & Delay
        \\ \hline
        \end{tabular}
  \caption{List of variables.}
%\LQ{Double Check the used symbols}}
\end{table}

\newpage
\section{Jaynes-Cumming nonlinearity readout and toy model}\label{sec:SIMethodTheory}

\subsection{Experimental observation}
\begin{figure}[ht!]
\centering
\includegraphics[scale=0.8]{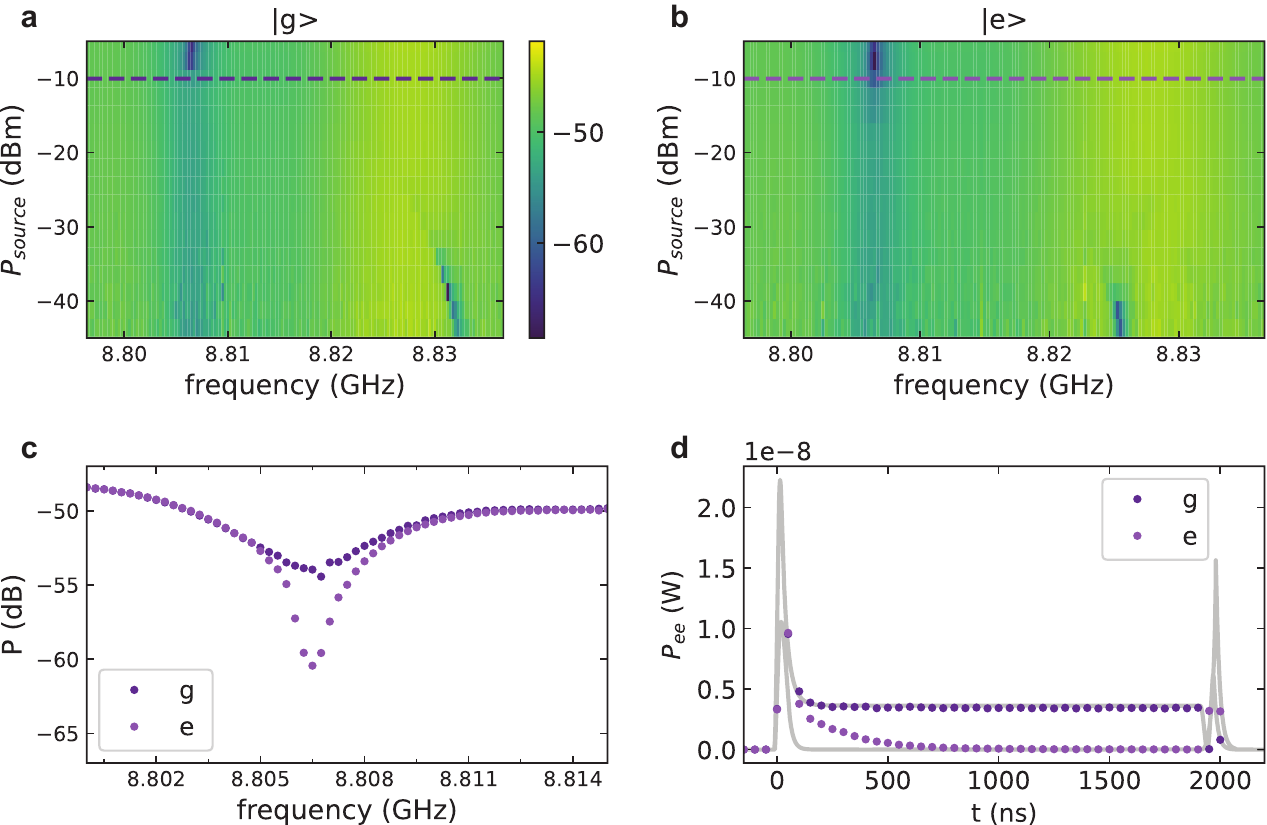}
\caption{
\textbf{Jaynes-Cumming readout. a}, Microwave reflection spectrum from power sweep of the cascaded cQED-EO cavities with the qubit in ground state, the electro-optic transducer cavity at $\omega_e/2\pi=8.806~\mathrm{GHz}$, and additionally the cQED cavity switched to $\omega_{bare} = \omega_e$ at high readout powers. 
\textbf{b}, Similar spectrum to \textbf{a} with the qubit prepared in the excited state.
The increased contrast due to the cQED cavity being resonant with the transducer cavity happens at slightly lower powers than for the ground state.
\textbf{c}, Reflection for both qubit states at the microwave readout power used in the main text for the Jaynes-Cumming readout. The contrast difference allows to detect the qubit states at high powers at the bare cQED cavity frequency. \textbf{d}, Averaged time trace for the all-optical readout, presented in Fig.2d. Unlike in the main text, we additionally show the modelled time trace for the qubit in the excited state with our cascaded cavity toy model without a qubit. While the ground state response and the steady state of the excited state reached after $\approx~1~\mu s$ can be sufficiently described, the dynamics of the qubit-cavity interaction at smaller times naturally deviates. 
}
\label{fig:SI1}
\end{figure}

In the dispersive limit, the nonlinearity from the Jaynes-Cumming interaction can be employed for a measurement of the qubit state with high signal-to-noise ratio~\cite{Reed2010a}. The measurement effectively makes use of the qubit-induced cavity anharmonicity (Lamb shift $\chi_0/2\pi~=~26~\mathrm{MHz}$) which depends on the cavity drive power for higher occupations and the dispersive shift of the cavity at low drive powers ($\chi/2\pi~=~6.6~\mathrm{MHz}$). The combination of both effects leads to the shift of the cavity resonance to its bare frequency $f_{bare}~=8.806~\mathrm{GHz}$ at different drive powers for different qubit states. This allows a qubit state-dependent readout with high SNR (cf. dashed lines in Fig.~\ref{fig:SI1}a and b). Theoretical descriptions of this behavior include either higher qubit levels~\cite{Boissonneault2010} or a semi-classical treatment for large cavity-qubit detunings and drive power-dependent anharmonicities~\cite{Bishop2010}. Both models led to qualitative agreement of the frequency spectrum with the experimental observation. Phenomenologically, the situation can be described by the cQED cavity resonance either being at its bare resonator frequency ($|e\rangle$) or being completely off-resonant ($|g\rangle$). In our scheme, the microwave cavity of the electro-optic transducer is tuned to $f_{bare}$. Thus, one observes either the transducer cavity alone or on resonance with the cQED cavity by means of cascaded cavities~(Fig.\ref{fig:SI1}c). Even though this omits the qubit-cavity interaction and can therefore not predict the temporal dynamics with the qubit being in $|e\rangle$, it is sufficient to model the system with independently calibrated parameters in steady state (Fig.\ref{fig:SI1}d).

%\subsection{Dispersive Microwave Readout}
%In the dispersive regime, the Hamiltonian of the qubit and cavity interaction takes the form,
%\begin{equation}
%{H}_c \approx \frac{1}{2} \hbar \omega_{\mathrm{q}} \hat{\sigma}_z+\hbar %\omega_{\mathrm{c}} \hat{a}_c^{\dagger} \hat{a}_c-\chi \hat{\sigma}_z %\hat{a}_c^{\dagger} \hat{a}_c
%\end{equation}
%with $\chi=g^2 / \Delta$ and $|\Delta|=\left|\omega_{c}-\omega_{q}\right| \gg g$.
%The qubit cavity frequency depends on the qubit state, i.e. $\omega_c \pm \chi$. The qubit state can be inferred from the microwave readout field from the qubit cavity
%$\hat{a}_{c,\mathrm{out}}=\hat{a}_{c,\mathrm{out}} - \sqrt{\kappa_{c,\mathrm{ex}}} \hat{a}_{c} (t)$.

%\begin{figure}
%    \centering
%    \includegraphics{MWReadout.pdf}
%    \caption{\textbf{Dispersive microwave readout}}
%    \label{figSI:mwreadout}
%\end{figure}
\subsection{Microwave readout model}
We model the microwave system consisting of the cQED cavity and the transducer microwave cavity as two cascaded cavities with reflective ports (coupling $\eta \kappa$) and intrinsic loss rates $(1-\eta) \kappa$. Specifically, the microwave readout field from the cQED cavity
$\hat{a}_{c,\mathrm{out}}(t)$ 
travels through superconducting cables of efficiency $\eta_{e,c}$ and delay $\tau$, and enters the microwave cavity of the electro-optical transducer with $\hat{a}_{e, \mathrm{in}}(t) = \eta_{e,c} \hat{a}_{c, \mathrm{out}}(t+\tau)$~\cite{C.W.Gardiner1992}. 

If we send the specific readout power to the cQED device, at which the Jaynes-Cumming nonlinearity allows to detect the qubit state~(Fig.~\ref{fig:SI1}c), the cQED cavity is either on-resonant with the transducer cavity ($\omega_e=\omega_c$, excited state) or completely reflective (ground state). We model the latter by a detuning $\chi_0$.

\begin{eqnarray}
		\dot{\hat{a}}_e=-\frac{\kappa_e}{2} \hat{a}_e
		+\sqrt{\eta_e \kappa_e} \eta_{e,c}\hat{a}_{c,\mathrm{out} },\\
		\dot{\hat{a}}_c=\left[-i \frac{\chi_0}{2} (\langle \hat{\sigma_z} \rangle +1)-\frac{\kappa_{c}}{2}\right] \hat{a}_c
		+\sqrt{\eta_c \kappa_{c}} \hat{a}_{c, \mathrm{in}}.     %\hat{a}_c^{\dagger} \hat{a}_c
\end{eqnarray}

We model the system using decoupled equations assuming the signal propagation is unidirectional. This is a reasonable assumption because of the microwave circulator between both cavities. 
%\textcolor{blue}{Check exact relation why the combined cavities are so well critically coupled}
We want to stress again that this is only a phenomenological model to describe the behavior of the system at the specific power chosen for the Jaynes-Cummings readout. 

\subsection{Microwave-optical readout model}
The microwave-optical readout of the qubit state is enabled by converting the microwave field from the cQED system into the optical domain using the electro-optic transducer. This allows for efficient conversion from microwave to optical fields. 
The electro-optic device is driven by a resonant optical pump pulse, with dynamics given by,
\begin{equation}
\dot{\bar{a}}_p  = \left(i \Delta_p-\frac{\kappa_p}{2}\right)\bar{a}_p
+\sqrt{\eta_p \kappa_p } \bar{a}_{p, \mathrm{in}},
\end{equation}
with $g(t) = \bar{a}_p(t) g_0$
the cavity enhanced electro-optical coupling rate.
The dynamics of the multi-mode electro-optic device can be described by the quantum Langevin equation,
\begin{eqnarray}
        \label{eq_EO1}
		\dot{\hat{a}}_e&=&-\frac{\kappa_e}{2} \hat{a}_e-i g\hat{a}_o -i g^* \hat{a}_s^{\dagger}
		+\sqrt{\eta_e \kappa_e} \hat{a}_{e, \mathrm{in}},\\
		%+ \sqrt{\left(1-\eta_e\right) \kappa_e} \delta\hat{a}_{e, 0}, \\
		\dot{\hat{a}}_o&=&\left(i \delta_o-\frac{\kappa_o}{2}\right) \hat{a}_o-i g \hat{a}_e , \\
		%+\sqrt{\eta_o \kappa_o} \delta\hat{a}_{o, \mathrm{in}} + \sqrt{\left(1- \eta_o\right) \kappa_o} \delta\hat{a}_{o, \mathrm{0}}, \\
		\dot{\hat{a}}_s&=&\left(i \delta_s-\frac{\kappa_s}{2}\right) \hat{a}_s-i g^* \hat{a}_e^{\dagger} - i J \hat{a}_{\mathrm{tm}},\\
        %+ \sqrt{\kappa_t} \delta\hat{a}_{t, \mathrm{vac}}, \\
		\dot{\hat{a}}_{\mathrm{tm}}&=&\left(i \delta_{\mathrm{tm}}-\frac{\kappa_{\mathrm{tm}}}{2}\right) \hat{a}_{\mathrm{tm}} - i J \hat{a}_s.
        %+ \sqrt{\kappa_{\mathrm{tm}}} \delta\hat{a}_{{\mathrm{tm}}, \mathrm{vac}},
        \label{eq_EO2}
\end{eqnarray}
where $a_{e,in}$ denotes again the output field from the cQED cavity. We note that, $J\ll\kappa_s$ in our device.
The qubit state is verified by the converted optical Stokes output field from the electro-optic device,
\begin{equation}
    \hat{a}_{o,\mathrm{out}} (t) = - \sqrt{\kappa_{o,\mathrm{ex}}} \hat{a}_o(t).
\end{equation}

%\begin{figure}
%    \centering
%    \includegraphics{MW-Optical Readout.pdf}
%        \caption{\textbf{Microwave-Optical Readout}}
%    \label{figSI:mwoptreadout}
%\end{figure}

\subsection{Full optical readout model}
The full optical readout of the superconducting qubit is realized by sending an optical signal together with an optical pulse to the electro-optic transducer. The converted microwave signal is used for qubit readout and reflected back to the electro-optic transducer. The reflected microwave field is subsequently converted again into the optical domain.
The dynamics of the electro-optic device from Eq.~\ref{eq_EO1}-\ref{eq_EO2} is now related to the cQED system by
%\begin{eqnarray}
%		\dot{\hat{a}}_e&=&-\frac{\kappa_e}{2} \hat{a}_e-i g\hat{a}_o -i g^* \hat{a}_s^{\dagger}
%		+\sqrt{\eta_e \kappa_e} \hat{a}_{e, \mathrm{in}},\\
%		\dot{\hat{a}}_o&=&\left(i \delta_o-\frac{\kappa_o}{2}\right) \hat{a}_o-i g \hat{a}_e 
%		+\sqrt{\eta_o \kappa_o} \hat{a}_{o, \mathrm{in}} , \\
%		\dot{\hat{a}}_s&=&\left(i \delta_s-\frac{\kappa_s}{2}\right) \hat{a}_s-i g^* %\hat{a}_e^{\dagger} - i J \hat{a}_{\mathrm{tm}},\\
%        %+ \sqrt{\kappa_t} \delta\hat{a}_{t, \mathrm{vac}}, \\
%		\dot{\hat{a}}_{\mathrm{tm}}&=&\left(i \delta_{\mathrm{tm}}-\frac{\kappa_{\mathrm{tm}}}{2}\right) \hat{a}_{\mathrm{tm}} - i J \hat{a}_s.
        %+ \sqrt{\kappa_{\mathrm{tm}}} \delta\hat{a}_{{\mathrm{tm}}, \mathrm{vac}},
%\end{eqnarray}
\begin{eqnarray}
\hat{a}_{e, \mathrm{in}}(t) & = \eta_{e,c} \hat{a}_{c, \mathrm{out}}(t),\\
\hat{a}_{c, \mathrm{in}}(t) & = \eta_{c,e}\hat{a}_{e, \mathrm{out}}(t).
\end{eqnarray}

The superconducting qubit state thus can be verified from the reflected optical anti-Stokes field,
\begin{equation}
    \hat{a}_{o,\mathrm{out}} (t) =\hat{a}_{o,\mathrm{in}} (t) - \sqrt{\kappa_{o,\mathrm{ex}}} \hat{a}_o(t).
\end{equation}

%\begin{figure}
%    \centering
%    \includegraphics{OpticalReadout.pdf}
%        \caption{\textbf{Optical Readout}}
%    \label{figSI:optreadout}
%\end{figure}

\section{Fabrication}\label{sec:SIFab}

\subsection{The cQED system}

For the fabrication of the transmon qubit, a 10 x 10~mm$^2$ high resistivity silicon chip was cleaned using an O$_2$ plasma asher followed by a buffered oxide etch (BOE) dip for 30s. The chip is further cleaned by sonicating in 50 $^{\circ}$C acetone for 10 mins and finally rinsed with isopropyl alcohol (IPA). The cleaned chip is covered with double layer of MMA/PMMA resist to pattern the qubit with two capacitor pads of area 330 x 550~$\mu$m$^2$ separated by a 400 $\mu$m distance along with a Josephson junction in the center of the gap between the two capacitors. The entire process is performed in a single e-beam lithography (EBL) step using the Dolan bridge process for an Al/AlO$_x$/Al junction of area 200 x 300~nm$^2$. For the metal deposition, an in-situ gentle argon ion milling is performed to clean any resist residues from the surface of the silicon. The Josephson junction is then fabricated by evaporating 60 nm aluminum followed by static oxidation in pure O$_2$ environment and then a 120 nm thickness aluminum layer is evaporated. The residual metal is lifted off using dimethyl sulfoxide (DMSO) at 80~$^{\circ}$C for 3 hours followed by an acetone and IPA rinse. Finally, the sample is covered with S1805 photo resist and UV tape as the 10 x 10~mm$^2$ chip is diced into three 10 x 2.5~mm$^2$ pieces with one qubit on each diced chip.

The transmon qubit is strongly coupled to a 23x15~mm rectangular waveguide cavity with rounded sidewalls made of 6061 Aluminum alloy. Indium seals the seam between the top and bottom part of the cavity. The fundamental mode can be tuned by a 8x8~mm Aluminum plate attached to an Aluminum rod inserted from an opening in the cavity housing. The central position of the tuner at the strongest field strength allows a large tuning range from $\approx~9~\mathrm{GHz}$ to $\approx~7~\mathrm{GHz}$ by the 5~mm range of motion of an Attocube ANPz101 piezomechanical nanopositioner.

The cQED system is mounted on a gold-plated oxygen-free high-conductivity copper holder and surrounded by a $\mu$-metal shield which protects it from static magnetic fields. Additionally, the magnetic shield of the lowest temperature stage of our Bluefors LD250 dilution refrigerator \cite{Bluefors2023} is painted by an infrared absorptive material.

%To shield the cQED system from electro-magnetic radiation and static fields, we use two shields at the mixing chamber as well as attenuators, circulators and filters in the microwave lines. Additionally,  Bluefors LD250 dilution refrigerator \cite{Bluefors2023} provides standard shielding. The qubit-cavity system is mounted on a copper holder and is surrounded by a $\mu$-metal shield which shields it from static magnetic fields. All devices mounted on the MXC are covered with a radiation shield painted in Stycast on its inside to enhance the blocking of RF and optical radiation. The low-pass filters and the eccosorbs block radiation which is outside of the frequency band we use, especially high-frequency radiation (IR frequencies). The attenuators in the mw lines diminish thermal noise from different stages. The circulators at the output attenuate the noise generated by the HEMT. Additional shielding from the IR radiation, we use for conversion in the electro-optic transducer, is not required.

\subsection{The electro-optic transceiver}
The electro-optic device comprises an optical cavity by means of a LiNbO$_3$ whispering gallery mode (WGM) resonator and a cylindrical 3D cavity made of pure Aluminum. The center of the cavity exhibits circular protrusions from the cavity top and bottom forming a narrow, ring-shaped gap. As the rings clamp the optical resonator close the optical modes confined at the rim and additionally strongly confine the electric field of the $m=1$ microwave resonance, the overlap between both fields is optimized. The microwave resonance can be tuned by roughly $500~\mathrm{MHz}$ inserting an aluminum cylinder into the cavity by another Attocube piezomechanical nanopositioner. GRIN lenses focus the optical input and output from the optical fiber on a diamond prism coupled to the WGM resonator. More details about fabrication and characterization can be found in \cite{hease2020}.

The device is identical to previous works \cite{hease2020,sahu2022, Sahu2023} but suffered from an internal linewidth increase from $2\pi\times7~\mathrm{MHz}$ to $\approx 2\pi\times40~\mathrm{MHz}$. Together with with a reduced mode matching factor for the coupling between the single mode optical fiber to the electro-optic device, this results in an almost two orders of magnitude decrease in conversion efficiency. We attribute the linewidth broadening to damages in the LiNbO$_3$ disc induced by the clamping rings due to thermal expansion during warmup. A soft Indium clamping in the center of the disc far from the optical modes resolved this issue in a recent device and the optical linewidth remains constant during cooldown and warmup. 

\section{Experiment}\label{sec:SISetup}

The experimental setup is described in Supplementary Fig.~\ref{fig:exp_setup} and the device parameters are listed in Supplementary Tab.~\ref{tab:parameters}.

\begin{figure*}[t]
    \centering
    \includegraphics[width=\linewidth]{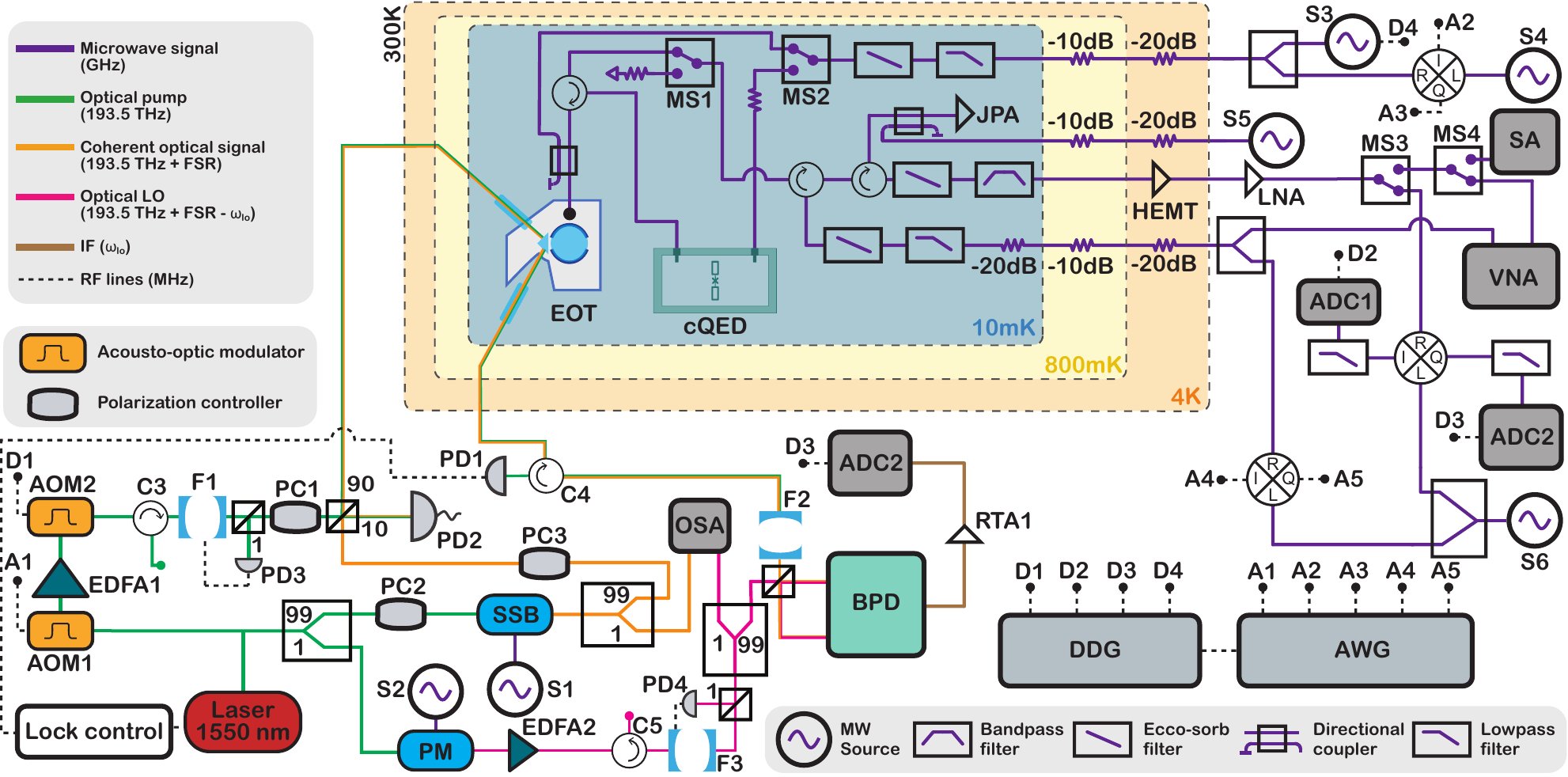}
    \caption{\textbf{Experimental setup}. \textbf{Optical setup (bottom left)} A telecom laser (Toptica DLC CTL 1500) gets split into the optical pump (green) and the optical signal (orange) and optical local oscillator (LO) for heterodyne detection (pink) respectively. The optical pump is amplified by an EDFA (Amonics AEDFA-PM-NS-200-10-23-M-FA) and pulse-shaped by two acousto-optical modulators (AOM, Fibre-Q T-M200-0.1C2J-3-F2P) with corresponding input signal from an arbitrary waveform generator (AWG, port A1, Spectrum Instrumentation M4i.6622-x8) and a digital delay/pulse generator (DDG, Stanford Research Systems DG645). An in-house filter cavity (F1) with analog PI-lock (PD3) removes the broadband noise from the EDFA amplification. The optical signal (orange) is generated by a single side-band modulator (SSB, Thorlabs DQPSK optical modulator) at the frequency of the Anti-Stokes mode, i.e. blue-shifted by 8.806~GHz. The optical LO (pink) is detuned from the laser frequency by a phase modulator (PM) and amplified. The PM's spurious modes are consequently cleaned by a tunable filter cavity (F3, MicronOptics FFPI) locked by software controlling a Peltier element. The optical output signal is cleaned from the reflected pump by another tunable filter (F2, MicronOptics FFPI) and finally combined with the optical LO for heterodyne detection on a balanced photodetector (Thorlabs PDB470). The downconverted signal is sent to an analog-digital converter (ADC2, port 2, AlazarTech ATS9870). \textbf{Microwave setup (top right)} An AWG (Quantum Machines OPX+) generates the qubit drive pulse via an IQ-mixer (ports A2, A3, Marki IQ-4509MXP). The pulse is sent either to the weakly coupled port of the qubit-cavity system (cQED system) or to the electro-optic transducer (EOT) for direct conversion measurements. The Quantum machines OPX+ also generates the qubit readout signal via an IQ-mixer (ports A4, A5, Marki IQ-0618MXP). The readout pulse is sent to the strongly coupled port of the cQED system. After being reflected there and at the EOT's microwave cavity, the pulse is routed via a reflective Josephson parametric amplifier \cite{Winkel2020}, a cryogenic HEMT amplifier(Low Noise Factory LNF-LNC6\_20C), and a room temperature low-noise amplifier (Agile AMT-A0067) to an IQ mixer. The downconverted signal from the IF port is sent to ADC2 (port 1) and the digitizer of Quantum Machines OPX+ (ADC1). Timing and synchronisation is controlled by the DDG. A vector network analyzer (VNA, Rohde and Schwartz ZVL13) can also be used to characterize the microwave setup. The microwave output or the signal from a 50~$\Omega$ termination (microwave switch MS1) may be sent to a spectrum analyzer (Rohde and Schwarz FSW26) for noise measurements. Further Acronyms: PC - polarization controller, PD - photodetector(Thorlbabs PDA50B2, PDA05CF2, PDA20CS2)}
    \label{fig:exp_setup}
\end{figure*}

%\subsection{Experimental parameters}

\begin{table}[H]
\caption{\textbf{Device parameters.}}
\centering
\begin{tabular}{|l|l|l|}
\hline Parameter & Symbol & Value \\
\hline Qubit frequency & $\omega_{\mathrm{q}}$ & $\omega_{\mathrm{q}} / 2 \pi=6.251 \mathrm{GHz}$ \\
\hline Qubit-cavity coupling & $g_{\mathrm{qc}}$ & $g_{\mathrm{qc}} / 2 \pi=326 \mathrm{MHz}$ \\
\hline Qubit anharmonicity & $\nu$ & $\nu / 2 \pi=201 \mathrm{MHz}$ \\
\hline Dispersive shift & $\chi$ & $\chi / 2 \pi=6.6 \mathrm{MHz}$ \\
\hline Cavity frequency & $\omega_{\mathrm{c}}$ & $\omega_{\mathrm{c}} / 2 \pi=8.806 \mathrm{GHz}$ \\
\hline Cavity linewidth & $\kappa_{\mathrm{c}}$ & $\kappa_{\mathrm{c}} / 2 \pi=1.4 \mathrm{MHz}$ \\
\hline Weak port coupling & $\kappa_{\mathrm{c}, \mathrm{w}}$ & $\kappa_{\mathrm{c}, \mathrm{w}} / 2 \pi=100 \mathrm{kHz}$ \\
\hline Cavity internal loss & $\kappa_{\mathrm{c}, \text { int }}$ & $\kappa_{\mathrm{c}, \text { int }} / 2 \pi=300 \mathrm{kHz}$ \\
\hline Qubit lifetime & $T_1$ & $T_1=40 \mu \mathrm{s}$ \\
\hline Ramsey time & $T_2$ & $T_2=1.5 \mu \mathrm{s}$ \\
\hline Optical cavity frequency & $\omega_{\mathrm{o}}$ & $\omega_{\mathrm{o}} / 2 \pi=193.4 \mathrm{THz}$ \\
\hline Optical cavity external coupling & $\kappa_{\mathrm{o}, \mathrm{ext}}$ & $\kappa_{\mathrm{o}, \text { ext }} / 2 \pi=44 \mathrm{MHz}$ \\
\hline Optical cavity linewidth & $\kappa_{\mathrm{o}}$ & $\kappa_{\mathrm{o}} / 2 \pi=81 \mathrm{MHz}$ \\
\hline EO Microwave cavity frequency & $\omega_{\mathrm{e}}$ & $\omega_{\mathrm{e}} / 2 \pi=8.806 \mathrm{GHz}$ \\
\hline EO Microwave cavity external coupling rate& $\kappa_{\mathrm{e}, \mathrm{ext}}$ & $\kappa_{\mathrm{e}, \mathrm{ext}} / 2 \pi=3.42 \mathrm{MHz}$ \\
\hline EO Microwave cavity linewidth & $\kappa_{\mathrm{e}}$ & $\kappa_{\mathrm{e}} / 2 \pi=9.69 \mathrm{MHz}$ \\
\hline Vacuum electro-optical coupling & $g_{\mathrm{eo}}$ & $g_{\mathrm{eo}} / 2 \pi=30 \mathrm{~Hz}$ \\
\hline
\end{tabular}
\label{tab:parameters}
\end{table}

%\section{Data Analysis}\label{sec:SIAnalysis}
\bibliography{Final_jf2v2}
\bibliographystyle{apsrev4-2}